\documentclass[12pt,a4paper,superscriptaddress,notitlepage]{revtex4-1}
\usepackage{pdfpages}
\makeatletter
\AtBeginDocument{\let\LS@rot\@undefined}
\makeatother
\usepackage{graphicx,hyperref} 
\usepackage{caption}
\usepackage{color}
\usepackage{ulem}
\usepackage{stackengine}
\usepackage{array}
\usepackage{multirow}
\usepackage{enumerate}
\usepackage{amsmath}
\usepackage{amssymb}
\usepackage{amsfonts}
\usepackage{ascmac}
\usepackage[nameinlink]{cleveref} 
\usepackage{url}
\usepackage{bm}
\usepackage[utf8]{inputenc}
\usepackage[T1]{fontenc}
\usepackage[scaled]{helvet}%
\usepackage{lmodern}
\crefname{equation}{Eq.\!}{Eqs.\!}
\crefname{figure}{Fig.\!}{Figs.\!}
\usepackage{blindtext}
\usepackage{microtype} 
\usepackage{lineno} 
\makeatletter
\newcommand*{\rom}[1]{\expandafter\@slowromancap\romannumeral #1@}
\makeatother
\usepackage{xcolor}
\hypersetup{
  colorlinks   = true, 
  urlcolor     = black, 
  linkcolor    = blue, 
  citecolor   = blue 
}
\def\ER{Erd\H{o}s-R\'{e}nyi\ }
\def\Exp{{\mathbb E}}

\newcolumntype{Y}{>{\centering\arraybackslash}X}

\def\Xmat{{\bm X}}
\def\RCA{R}

\def\Xmatrand{\tilde {\bm X}}

\def\xvec{{\bm x}}

\def\EBIC{\text{EBIC}}
\def\qCov{{{\bm W}}}
\def\Pre{{{\bm \Theta}}}
\def\nPre{{{\bm \Theta}^{\text{null}}}}
\def\qPre{{{\bm W}}}
\def\qpre#1{{W_{#1}}}
\def\qCovtemp{\qCov}

\def\qcov#1{{W_{#1}}}
\def\Cov{{\bm C}}

\def\nCov{{\bm C}^{\text{null}}}
\def\sCov{{\bm C}^{\text{sample}}}
\def\sCovrand{\tilde {\bm C}^{\text{sample}}}
\def\scovrand#1{\tilde C_{#1}^{\text{sample}}}
\def\scov#1{C_{#1} ^{\text{sample}}}
\def\cov{{C}}
\def\ncov#1{{C_{#1}^{\text{null}}}}
\def\trace{\text{tr}}

\definecolor{myblue}{RGB}{33,150,243}
\definecolor{mygreen}{RGB}{76, 175, 80}
\definecolor{purple}{RGB}{170, 0, 255}

\begin{document}
\title{Constructing networks by filtering correlation matrices: A null model approach}
\author{Sadamori Kojaku}
\author{Naoki Masuda}
\email{naoki.masuda@bristol.ac.uk}
\affiliation{
Department of Engineering Mathematics,
Merchant Venturers Building, University of Bristol,
Woodland Road, Clifton, Bristol BS8 1UB, United Kingdom
}

\begin{abstract}
Network analysis has been applied to various correlation matrix data.
Thresholding on the value of the pairwise correlation is probably the most straightforward and common method to create a network from a correlation matrix. 
However, there have been criticisms on this thresholding approach such as an inability to filter out spurious correlations, which have led to proposals of alternative methods to overcome some of the problems. 
We propose a method to create networks from correlation matrices based on optimisation with regularization, where we lay an edge between each pair of nodes if and only if the edge is unexpected from a null model. 
The proposed algorithm is advantageous in that it can be combined with different types of null models.
Moreover, the algorithm can select the most plausible null model from a set of candidate null models using a model selection criterion. 
For three economic data sets, we find that the configuration model for correlation matrices is often preferred to standard null models.
For country-level product export data, the present method better predicts main products exported from countries than sample correlation matrices do.
\end{abstract}

\maketitle
\section{Introduction}

Many networks have been constructed from correlation matrices.
For instance, asset graphs are networks in which a node represents a stock of a company and an edge between a pair of nodes indicates strong correlations between two stock prices \cite{Onnela2004,Tse2010}.
A variety of tools for network analysis, such as centralities, network motifs and community structure, can be used for studying properties of correlation networks \cite{Newman2018}. 
Network analysis may provide information that is not revealed by other analysis methods for correlation matrices such as principal component analysis and factor analysis.
Network analysis on correlation data is commonly accepted across various domains \cite{Onnela2004,Rubinov2010,VanWijk2010,Zanin2012,DeVicoFallani2014,Kose2001}.
 
The present paper proposes a new method for constructing networks from correlation matrices.  
There exist various methods for generating sparse networks from correlation matrices such as those based on the minimum spanning tree \cite{Mantegna1999} and its variant \cite{Tumminello2005}.
Perhaps the most widely used technique is thresholding, i.e., retaining pairs of variables as edges if and only if the correlation or its absolute value is larger than a threshold value \cite{Kose2001,Onnela2004,Rubinov2010,VanWijk2010,Zanin2012,DeVicoFallani2014}. 
Because the structure of the generated networks may be sensitive to the threshold value, 
a variety of criteria for choosing the threshold value have been proposed \cite{Onnela2004,Rubinov2010,VanWijk2010,Zanin2012,DeVicoFallani2017}. 
An alternative is to analyse a collection of networks generated with different threshold values \cite{Zanin2012}.

The thresholding method is problematic when large correlations do not imply dyadic relationships between nodes \cite{Jackson1991,Vigen2015}. 
For example, in a time series of stock prices, global economic trends (e.g., recession and inflation) simultaneously affect different stock prices, which can lead to a large correlation between various pairs of stocks \cite{Bouchaud2003,MacMahon2015}.
Additionally, the correlation between two nodes may be accounted for by other nodes.
A major instance of this phenomenon is that, if node $v_1$ is strongly correlated with nodes $v_2$ and $v_3$, then $v_2$ and $v_3$ would be correlated even if there is no direct relationship between them \cite{Marrelec2006,Zalesky2012,Masuda2018b}.
For example, the murder rate (corresponding to node $v_2$) is positively correlated with the ice cream sales (node $v_3$), which is accounted for by the fluctuations in the temperature (node $v_1$), i.e.,  people are more likely to interact with others and purchase ice creams when it is hot \cite{Vigen2015}.

The sparse covariance estimation \cite{Bien2011} and graphical Lasso \cite{Banerjee2008,Friedman2008,Fan2009,Foygel2010,VanBorkulo2014} were previously proposed for filtering out spurious correlations.
These methods consider the correlation between nodes to be spurious if the correlation is accounted for by random fluctuations under a white noise null model, which assumes that observations at all nodes are independent of each other.
However, nodes tend to be correlated with each other in empirical data owing to trivial factors (e.g., global trend), which calls for different null models that emulate different types of spurious correlations \cite{Laloux1999,Plerou1999,Hirschberger2007,MacMahon2015,Masuda2018}.

To incorporate such null models into network inference, we present an algorithm named the Scola, standing for Sparse network construction from COrrelational data with LAsso.
The Scola places edges between node pairs if the correlations are not accounted for by a null model of choice.
A main advantage of the Scola is that it leaves the choice of null models to users, enabling them to filter out different types of trivial relationships between nodes. 
Furthermore, the Scola can select the most plausible null model for the given data among a set of null models using a model selection framework.
A Python code for the Scola is available at \cite{scola-code}

\section{Methods}

\subsection{Construction of networks from correlation matrices}
\label{sec:lasso}

Consider $N$ variables, which we refer to as nodes. 
We aim to construct a network on $N$ nodes, in which edges indicate the correlations that are not attributed to some trivial properties of the system.
Let $\Cov=(\cov_{ij})$ be the $N \times N$ correlation matrix, where $\cov_{ij}$ is the correlation between nodes $i$ and $j$, i.e., $-1\leq \cov_{ij} \leq 1$ and $\cov_{ii}=1$ for $1\leq i \leq N$.
We write $\Cov$ as 
\begin{align}
	\label{eq:cov}
	\Cov = \nCov + \qCov.
\end{align}
Matrix $\nCov = \left(\ncov{ij}\right)$ is an $N \times N$ correlation matrix, where $\ncov{ij}$ is the mean value of the correlation between nodes $i$ and $j$ under a null model.
For example, if every node is independent of each other under the null model, then one sets $\nCov ={\bm I}$, where ${\bm I}$ is the $N \times N$ identity matrix. 
We introduce three null models in Section~\ref{sec:null-models}. 
Matrix $\qCov = \left(\qcov{ij}\right)$ is an $N \times N$ matrix representing the deviation from the null model.
We place an edge between nodes $i$ and $j$ (i.e., $W_{ij}\neq 0$) if and only if the correlation $C_{ij}$ is sufficiently different from that for the null model.
We note that edges are undirected and may have positive or negative weights. 
The network is assumed not to have self-loops (i.e., $\qcov{ii} =0$) because the diagonal elements of $\nCov$ and $\sCov$ are always equal to one.

We estimate $\qCov$ from data as follows.
Assume that we have $L$ samples of data observed at the $N$ nodes, based on which the correlation matrix is calculated.
Denote by $\Xmat=(x_{\ell i})$ the $L\times N$ matrix, in which $x_{\ell i}$ is the value observed at node $i$ in the $\ell$th sample.
Let $\xvec_{\ell }=\left[x_{\ell 1}, x_{\ell 2}, \ldots, x_{\ell N}\right]$ be the $\ell$th sample. 
We assume that each sample $\xvec_{\ell}$ is independently and identically distributed according to an $N$-dimensional multivariate Gaussian distribution with mean zero. 
For mathematical convenience, we assume that $\Xmat$ is preprocessed such that the average and the variance of $x_{\ell i}$ over the $L$ samples are zero and one respectively, i.e., $\sum_{\ell =1}^L x_{\ell i}/L = 0$ and $\sum_{\ell =1}^L x_{\ell i}^2 /L = 1$ for $1 \leq i \leq N$.

Our goal is to find $\qCov$ that maximises the likelihood of $\Xmat$, i.e., 
\begin{align}
	\label{eq:likelihood}
	P\left(\Xmat \vert \Cov \right) &\equiv \prod_{\ell =1}^L \frac{1}{(2\pi)^{N/2}\det\left(\Cov\right)^{1/2}} \exp\left( -\frac{1}{2} \xvec_{\ell } \Cov^{-1}  \xvec_{\ell } ^\top \right),
\end{align}
where $\top$ is the transposition.
It should be noted that $\Cov$ may not be equal to the sample Pearson correlation matrix, denoted by $\sCov\equiv \Xmat^{\top} \Xmat/L$ \cite{Kollo2006}, if $L$ is finite. 
The log likelihood is given by  
\begin{align}
	\label{eq:loglikelihood}
	\ln P\left(\Xmat \vert \Cov \right) &= -\frac{L}{2}\ln\det\left(\Cov\right) - \frac{1}{2}\sum_{\ell =1}^L \xvec_{\ell }  \Cov^{-1} \xvec_{\ell} ^{\top}  - \frac{NL}{2} \ln(2\pi).
\end{align}
Using $\sum_{\ell =1}^L\xvec_{\ell }  \Cov^{-1}\xvec_{\ell } ^{\top}= \trace\left(\Xmat ^{\top}\Xmat \Cov^{-1}\right)$, one obtains
\begin{align}
	\label{eq:logp}
	\ln P\left(\Xmat \vert \Cov \right) & = - \frac{L}{2}\ln\det\left(\Cov\right) - \frac{L}{2}\trace\left(\Cov^{\text{sample}}\Cov^{-1} \right)  - \frac{NL}{2} \ln(2\pi). 
\end{align}
Substitution of Eq.~\eqref{eq:cov} into Eq.~\eqref{eq:logp} leads to the log likelihood of $\qCov$ as follows. 
\begin{align}
	{\cal L}\left(\qCov\right) & \equiv -\frac{L}{2}\ln\det\left(\nCov + \qCov\right) - \frac{L}{2}\trace\left[\Cov^{\text{sample}}\left(\Cov^{\text{null}}  + \qCov\right)^{-1} \right] - \frac{NL}{2} \ln(2\pi). 
\end{align}
The log likelihood ${\cal L}$ is a concave function with respect to $\qCov$.
Therefore, one obtains the maximiser of ${\cal L}$, denoted by $\qCov^{\text{MLE}}$, by solving $\partial {\cal L}/\partial \qCov = 0$, i.e.,  
\begin{align}
	\label{eq:mle}
	\qCov^{\text{MLE}} \equiv \sCov - \nCov.
\end{align}

In practice, $\qCov^{\text{MLE}}$ overfits the given data, leading to a network with many spurious edges.
This is because the number of samples, $L$, is often smaller than the number of elements in $\qCov^{\text{MLE}}$, i.e.,  $L < N(N-1)/2$, as is the case for the estimation of correlation matrices and precision matrices \cite{Banerjee2008,Bien2011}. 
To prevent overfitting, we impose penalties on the number of non-zero elements in $\qCov$ using the Lasso \cite{Tibshirani1996}. 
The Lasso is commonly used for regression analysis to obtain a model with a small number of non-zero regression coefficients. 
The Lasso is also used for estimating sparse covariance matrices (i.e., $\Cov$) \cite{Bien2011} and sparse precision matrices (i.e., $\Cov^{-1}$) \cite{Banerjee2008} with a small number of samples. (We discuss these methods in Section~\ref{sec:discussion}.) 
Here we apply the Lasso to obtain a sparse $\qCov$. 
Specifically, we maximise penalized likelihood function
\begin{align}
	\label{eq:plikelihood}
	\hat {\cal L}\left(\qCov \vert {\bm \lambda} \right) \equiv {\cal L}\left(\qCov\right)
			- \frac{L}{2}\sum_{i=1}^N \sum_{j=1}^{i-1} \lambda_{ij} |\qcov{ij}|, 
\end{align}
where $\lambda_{ij} \geq 0$ is the Lasso penalty for $\qcov{ij}$.
Large values of $\lambda_{ij}$ yield sparse $\qCov$.
Because $\hat {\cal L}$  is not concave with respect to $\qCov$, one cannot analytically find the maximum of $\hat {\cal L}$. 
Therefore, we numerically maximise $\hat {\cal L}$ using an extension of a previous algorithm \cite{Bien2011}, which is described in Section~\ref{sec:algorithm}. 

The penalized likelihood $\hat {\cal L}$ contains $N(N-1)/2$ Lasso penalty parameters, $\lambda_{ij}$.
A simple choice is to use the same value for all $\lambda_{ij}$.
However, this method is problematic \cite{Fan2001,Zou2006}.
If one imposes the same penalty to all node pairs, 
one tends to obtain either a sparse network with small edge weights or a dense network with large edge weights. 
However, sparse networks with large edge weights or dense networks with small edge weights may yield a larger likelihood.
A remedy for this problem is the adaptive Lasso \cite{Zou2006}, which sets 
\begin{align}
	\label{eq:alasso}
	\lambda_{ij} = \overline \lambda |W^{\text{MLE}} _{ij}|^{-\gamma}, 
\end{align}
where $\overline \lambda \geq 0$ and $\gamma > 0$ are hyperparameters.
With the adaptive Lasso, a small penalty is imposed on a pair of nodes $i$ and $j$ if $W^{\text{MLE}} _{ij}$ is far from zero, allowing the edge to have a large weight.
If one has sufficiently many samples, the adaptive Lasso correctly identifies zero and non-zero regression coefficients (i.e., $\qcov{ij}$ in our case) given an appropriate $\overline \lambda$ value and any positive $\gamma$ value \cite{Zou2006}.
The estimated $\qcov{ij}$ values did not much depend on $\gamma$ in our numerical experiments.
Therefore, we set $\gamma=2$, which is a typical value \cite{Zou2006,Fan2009,Hui2015}.
Hyperparameter $\overline \lambda$ controls the number of edges in the network (i.e., the number of nonzero elements in $\qCov$). A large $\overline \lambda$ yields sparse networks.  
We describe how to choose $\overline \lambda$ in Section~\ref{sec:model-selection}.

\subsection{Null models for correlation matrices}
\label{sec:null-models}

The Scola accepts various null correlation matrices, i.e., $\nCov$.
Nevertheless, we mainly focus on the configuration model for correlation matrices \cite{Masuda2018}.
Although we also examined two other null models in numerical experiments, the configuration model was mostly chosen in the model selection (Section~\ref{sec:model-selection-result}).

Our configuration model is based on the principle of maximum entropy.
With this method, one determines the probability distribution of data, denoted by $P(\tilde \Xmat)$, by maximising the following Shannon entropy given by
\begin{align}
	\label{eq:H}
	- \int P(\Xmatrand)\ln P(\Xmatrand){\rm d}\Xmatrand, 
\end{align}
with respect to $P(\Xmatrand)$ under constraints, where $\Xmatrand=(\tilde x_{\ell i})$ is an $L \times N$ matrix such that $\tilde x_{\ell i}$ is the value at node $i$ in the $\ell$th sample.
Let $\sCovrand \equiv \Xmatrand^\top \Xmatrand/L$ be the sample covariance matrix for $\tilde \Xmat$. 
In the configuration model, we impose that each node has the same expected variance as that for the original data, i.e., 
\begin{align}
	\label{eq:Hconst1}
	\int \scovrand{ii} P(\Xmatrand) {\rm d}\Xmatrand = \scov{ii},\ \text{for $1 \leq i \leq N$}.
\end{align}
We also impose that the row sum (equivalently, the column sum) of $\sCov$ is preserved, i.e.,  
\begin{align}
	\label{eq:Hconst2}
	\int \sum_{j=1}^N \scovrand{ij} P(\Xmatrand) {\rm d}\Xmatrand = \sum_{j=1}^N \scov{ij},\ \text{for $1 \leq i \leq N$}.
\end{align}
Equation~\eqref{eq:Hconst2} is analogous to the case of the configuration model for networks, which by definition preserves the row sum of the adjacency matrix of a network, or equivalently the degree of each node.
We note that the $i$th row sum of $\sCov$ is proportional to the correlation between the observation at the $i$th node and the average of the observations over all nodes \cite{MacMahon2015}.

Denote by $\Cov^{\text{con}}$ the expectation of $\sCovrand$.
We compute $\Cov^{\text{con}}$, which we use as the null correlation matrix (i.e., $\nCov$), as follows.
In Ref.~\cite{Masuda2018}, we showed that $P(\Xmatrand)$ is a multivariate Gaussian distribution with mean zero.
Under $P(\Xmatrand)$, $\Cov^{\text{con}}$ is equal to the variance parameter of $P(\Xmatrand)$ \cite{Gupta2000,Kollo2006,Masuda2018}.
By substituting Eq.~\eqref{eq:likelihood} into Eqs.~\eqref{eq:H}, \eqref{eq:Hconst1} and \eqref{eq:Hconst2}, we rewrite the maximisation problem as   
\begin{align}
	\label{eq:H_}
	\max_{\Cov^{\text{con}}} \ln\det\left(\Cov^{\text{con}}\right)
\end{align}
subject to
\begin{align}
	\cov^{\text{con}} _{ii} = \scov{ii}\ \text{and}\ \sum_{j=1}^N \cov^{\text{con}} _{ij} = \sum_{j=1}^N\scov{ij},\ \text{for $1 \leq i \leq N$}. \label{eq:Hconst_}
\end{align}
Equation~\eqref{eq:H_} is concave with respect to $\Cov^{\text{con}}$. 
Moreover, the feasible region defined by Eq.~\eqref{eq:Hconst_} is a convex set.
Therefore, the maximisation problem is a convex problem such that one can efficiently find the global optimum.  
We compute $\Cov^{\text{con}}$ using an in-house Python program available at  \cite{dmccalg}.

In addition to the configuration model, we consider two other null models.
The first model is the white noise model, in which the signal at each node is independent of each other and has the same variance with that in the original data. 
The null correlation matrix for the white noise model, denoted by $\Cov^{\text{WN}}$, is given by $\cov^{\text{WN}} _{ij}=0$ for $i\neq j$ and $\cov^{\text{WN}} _{ii}=1$ for $1 \leq i \leq N$.
The white noise model is often used in the analysis of correlation networks \cite{Fisher1915,MacMahon2015}.

Another null model is the Hirschberger-Qi-Steuer (HQS) model \cite{Hirschberger2007}, in which each node pair is assumed to be correlated to the same extent as expectation.
As is the case for the configuration model for correlation matrices, the original HQS model provides a probability distribution of covariance matrices. 
The HQS model preserves the variance of the signal at each node averaged over all the ndoes as expectation.
Moreover, the HQS model preserves the average and variance of the correlation values over different pairs of nodes in the original correlation matrix as expectation.  
The HQS model is analogous to the \ER random graph for networks, in which each pair of nodes is adjacent with the same probability \cite{Masuda2018}.
We use the expectation of the correlation matrix generated by the HQS model, denoted by $\Cov^{\text{HQS}}$. 
We obtain $\cov_{ij} ^{\text{HQS}} = \left[N(N-1)/2\right]^{-1} \sum_{i^{\prime}=1}^N \sum_{j^{\prime}=1}^{i^{\prime}-1} \scov{i^{\prime}j^{\prime}}$ for $i\neq j$ and $\cov_{ii} ^{\text{HQS}}=1$ \cite{Hirschberger2007}.

\subsection{Model selection}
\label{sec:model-selection}

We determine the value of $\overline \lambda$ based on a model selection criterion.  
Commonly used criteria, Akaike Information Criterion (AIC) and Bayesian Information Criterion (BIC), favour an excessively rich model if the model has many parameters relative to the number of samples \cite{Chen2008}.
As discussed in Section \ref{sec:lasso}, this is often the case for the estimation of correlation matrices \cite{Banerjee2008,Foygel2010,Bien2011}. 
 
We use the extended BIC (EBIC) to circumvent this problem \cite{Chen2008,Foygel2010}.
Let $\hat \Cov^{(\overline \lambda)}$ be the estimated correlation matrix at $\overline \lambda$, i.e., $\hat \Cov^{(\overline \lambda)} = \nCov + \qCov^{(\overline \lambda)}$, where 
$\qCov^{(\overline \lambda)}$ is the network one estimates by maximising $\hat {\cal L}(\qCov|\overline \lambda)$. 
We adopt $\hat \Cov ^{(\overline \lambda)}$ that minimises 
\begin{align}
	\label{eq:EBIC}
    \EBIC_{\beta} = - 2\ln P\left(\Xmat \vert \hat \Cov^{(\overline \lambda)}\right) + \left(M + K \right) \ln L + 4\beta \left(M + K \right) \ln N, 
\end{align}
with respect to $\overline \lambda$. 
In Eq.~\eqref{eq:EBIC}, $M$ is the number of edges in the network (i.e., the half of the number of nonzero elements in $\qCov$), $K$ is the number of parameters of the null model and $\beta$ is a parameter.
The white noise model introduced in Section~\ref{sec:null-models} does not have parameters. 
Therefore, $K=0$.
The HQS model has $K=1$ parameter, i.e., the average of the off-diagonal elements.
The configuration model yields $K=N$, i.e., the row sum of each node.
It should be noted that we compute the number of parameters by exploiting the fact that $\sCov$ is a correlation matrix, i.e., the diagonal entries of $\sCov$ are always equal to one.

Parameter $\beta \in [0,1]$ determines the prior distribution for a Bayesian inference.  
The prior distribution affects the sparsity of networks; a large $\beta$ value would yield sparse networks.
Nevertheless, the effect of the prior distribution on the EBIC value diminishes when the number of samples (i.e., $L$) increases.
We adopt a typical value, i.e., $\beta = 0.5$, which provided reasonable results for linear regressions and the estimation of precision matrices \cite{Chen2008,Foygel2010,VanBorkulo2014}.
We adopt the golden-section search method to find the $\overline \lambda$ value that yields the minimum EBIC value \cite{Press2007} (Appendix \ref{sec:golden-section-search}). 

In addition to selecting the $\overline \lambda$ value, the EBIC can be used for selecting a null model among different types of null models.
In this case, we compute $\qCov$ for each null model with the $\overline \lambda$ value determined by the golden-section search method. 
Then, we select the pair of $\nCov$ and $\qCov$ that minimises the EBIC value.

\subsection{Maximising the penalized likelihood}
\label{sec:algorithm}

To maximise $\hat {\cal L}$ in terms of $\qCov$, we use the minorise-maximise (MM) algorithm \cite{Bien2011}. 
Although the MM algorithm may not find the global maximum, it converges to a local optimum. 
The MM algorithm starts from initial guess $\qCov = {\bm 0}$ and iterates rounds of the following minorisation step and the maximisation step. 

In the minorisation step, we approximate $\hat {\cal L}$ around the current estimate $\overline \qCov$ by \cite{Bien2011}
\begin{align}
	\label{eq:fx2}
	F_{\overline \qCov}\left(\qCovtemp\right) 
		\equiv -\frac{L}{2}\ln\det\left(\Cov^{\text{null}} + \overline \qCov\right)
		- \frac{L}{2}\trace\left[\left(\Cov^{\text{null}} + \overline \qCov \right)^{-1}\left(\Cov^{\text{null}} + \qCovtemp \right) \right]  \nonumber \\
		- \frac{L}{2}\trace\left[ \Cov^{\text{sample}}\left(\Cov^{\text{null}} + \qCovtemp \right)^{-1}\right]
		- \frac{NL}{2} \ln(2\pi) 
		- \frac{L}{2}\sum_{i=1} ^N \sum_{j=1}^{i-1} \lambda_{ij} |\qcov{ij}| + \frac{NL}{2}.
\end{align}
It should be noted that $F_{\overline \qCov}$ is a minoriser of $\hat {\cal L}$ satisfying $F_{\overline \qCov}(\overline \qCov) = \hat {\cal L}(\overline \qCov)$ and $F_{\overline \qCov}(\qCovtemp) \leq \hat {\cal L}(\qCovtemp)$ for all $\qCovtemp$ provided that $\nCov + \qCov$ is positive definite \cite{Bien2011}.
This property of $F_{\overline \qCov}$ ensures that maximising $F_{\overline \qCov}$ yields a $\hat {\cal L}$ value larger than or equal to  $\hat {\cal L}(\overline \qCov)$.

In the maximisation step, we seek the maximiser of $F_{\overline \qCov}$.
Function $F_{\overline \qCov}$ is a concave function, which allows us to find the global maximum with a standard gradient descent rule for the Lasso. 
Specifically, starting from $\widetilde \qCov^{(0)} = \overline \qCov$, we iterate the following update rule until convergence:
\begin{align}
\label{eq:update}
\widetilde \qCovtemp^{(k+1)} &= {\cal S}_{\epsilon {\bm \lambda}}\left( \widetilde \qCovtemp^{(k)} - \epsilon \left[ \left(\nCov + \overline \qCov \right)^{-1} -\left( \nCov + \widetilde \qCovtemp^{(k)} \right)^{-1} \sCov \left( \nCov + \widetilde \qCovtemp^{(k)} \right)^{-1} \right]\right),
\end{align}
where $\widetilde \qCovtemp^{(k)}$ is the tentative solution at the $k$th iteration, $\epsilon$ is the learning rate, and ${\cal S}$ is the element-wise soft threshold operator given by 
\begin{align}
	\left({\cal S}_{\epsilon {\bm \lambda}}(Z)\right)_{ij} = \left\{ 
	\begin{array}{ll}
		Z_{ij} - \epsilon \lambda_{ij} & (\epsilon \lambda_{ij} < Z_{ij} ), \\
		0 & (-\epsilon \lambda_{ij} \leq Z_{ij} \leq \epsilon \lambda_{ij}), \\
		Z_{ij} + \epsilon \lambda_{ij} & (Z_{ij} < -\epsilon \lambda_{ij}). 
	\end{array}
\right.
\end{align}
If  $\hat {\cal L}(\widetilde \qCovtemp^{(k_{\text{final}})}) > \hat {\cal L}(\overline \qCov)$, where $k_{\text{final}}$ is the iteration number after a sufficient convergence, we set $\overline \qCov = \widetilde \qCovtemp^{(k_{\text{final}})}$. Then, we perform another round of a minorisation step and a maximisation step.
Otherwise, the algorithm finishes.

The learning rate $\epsilon$ mainly affects the speed of the convergence of the iterations in the maximisation step.
We set $\epsilon$ using the ADAptive Moment estimation (ADAM), a gradient descent algorithm used in various machine learning algorithms \cite{Kingma2014}. 
ADAM adapts the learning rate at each update (Eq.~\eqref{eq:update}) based on the current and past gradients.
%

The MM algorithm requires computational time of ${\cal O}(N^3)$ because Eq.~\eqref{eq:update} involves the inversion of $N \times N$ matrices.
In our numerical experiments (Section~\ref{sec:casestudy}), the entirely of the Scola consumed approximately three hours on average for a network with $N=$1,930 nodes with 16 threads running on the Intel 2.6 GHz Sandy Bridge processors with 4GB memory.

\section{Numerical results}
\label{sec:casestudy}

\subsection{Prediction of country-level product exports}
\label{sec:product-space}

The product space is a network of products, where a pair of products is defined to be adjacent if both of them have a large share in the export volume of the same country \cite{Hidalgo2007,Hartmann2017}.
Here we construct the product space with the Scola and use it for predicting product exports from countries.

We use the data set provided by the Observatory of Economic Complexity \cite{product-space}, which contains annual export volumes of $N_p = 988$ products from $N_c = 249$ countries between 1962 and 2014.
The data set also contains the Standard International Trade Classification (SITC) code for each product, which indicates the product type.
We quantify the level of sophistication of the product types using the PRODY index \cite{Hausmann2007}.
We compute the PRODY index in 1991 for each product, where a product with a large PRODY index is considered to be sophisticated.
Then, we average the PRODY index over the products of the same type.
The product types in descending order of the PRODY index are as follows: 
``Machinery \& transport equipment'', ``Chemicals'', ``Miscellaneous manufactured articles'', ``Manufactured goods by material'', ``Miscellaneous'', 
``Mineral fuels, lubricants \& related materials'', ``Beverages \& tobacco'', ``Animal \& vegetable oils, fats \& waxes'', ``Food \& live animals'' and ``Raw materials''. 

In previous studies, the product space was constructed based on the so-called revealed comparative advantage (RCA) defined by
\begin{align}
\RCA _{cp}^{(t)} \equiv 
	\frac{
		V_{cp} ^{(t)} / \sum_{p'=1} ^{N_p} V _{cp'} ^{(t)}
	}{
		\sum_{c'=1}^{N_c} V_{c'p} ^{(t)} / \sum_{c'=1}^{N_c} \sum_{p'=1}^{N_p}V_{c'p'} ^{(t)}
	}, 
\end{align}
where $V_{cp}^{(t)}$ is the total export volume of product $p$ from country $c$ in year $t$.
Then, it was assumed that two products $p$ and $p'$ were adjacent in the product space if and only if $\RCA _{cp}^{(t)}, \RCA _{cp'}^{(t)}>1$ for at least one country $c$ \cite{Hidalgo2007}.

In contrast to this approach, we construct the product space as follows.
First, countries may export different products in different years.
To mitigate the effect of temporal changes, we split the data into two halves, i.e., those between $T_{\text{s}}=1972$ and $T_{\text{f}}=1991$, and those between $T_{\text{s}}=1992$ and $T_{\text{f}}=2011$.
 
Second, for each time window, we apply the Box-Cox transformation \cite{Box1964} to each $\RCA _{cp}^{(t)}$ to make the distribution closer to a standard normal distribution (Appendix~\ref{sec:box-cox}).
This preprocessing is crucial because the Scola assumes that the given data, $\Xmat$, is distributed according to a multivariate Gaussian distribution.
We note that the sample average and variance of the transformed RCA values based on the $N_p$ products are equal to zero and one, respectively.

Third, we define a sample $\xvec_{(c,t)}$ for each combination of country $c$ and year $t$ by 
\begin{align}
	\xvec_{(c,t)} \equiv \left[ \overline \RCA_{c1}^{(t)}, \ldots, \overline \RCA_{cN_p}^{(t)}\right],
\end{align}
where $\overline \RCA _{cp} ^{(t)}$ is the transformed value of $\RCA _{cp} ^{(t)}$.

Fourth, we compute a sample Pearson correlation matrix for concatenated samples $\left[\xvec_{(c,t)},\xvec_{(c,t+10)}\right]$. 
Specifically, for each of the two time windows, we compute $\sCov$ by
\begin{align}
	\sCov  &= \frac{1}{N_c (T_{\text{f}}-T_{\text{s}}-9)} \sum_{c=1}^{N_c}\sum_{t=T_{\text{s}}}^{T_{\text{f}}-10} \left[\xvec_{(c,t)},\xvec_{(c,t+10)}\right]^{\top} \left[\xvec_{(c,t)},\xvec_{(c,t+10)}\right] \nonumber \\ 
	       &=\left[
	\begin{array}{cc}
		\sCov _{0, 0} & \sCov _{0, +10} \\
		(\sCov _{0, +10}) ^\top & \sCov _{+10, +10} 
	\end{array} \right].
	\label{eq:scov-block}
\end{align}
In Eq.~\eqref{eq:scov-block}, $\sCov_{0,0}$ and $\sCov_{+10,+10}$ are the correlation matrices for the products within the same year.
The off-diagonal block $\sCov_{0,+10}$ contains the correlations between the products with a time lag of ten years. 

Fifth, we generate networks by applying either the thresholding method or the Scola to $\sCov$. 
For the Scola, we adopt the configuration model as the null model.
For the thresholding method, we set the threshold value such that the number of edges in the network is equal to that in the network generated by the Scola.
We set the weight of each edge to one for the network generated by the thresholding method.

\begin{figure}
\centering
    \includegraphics[width=0.77\hsize]{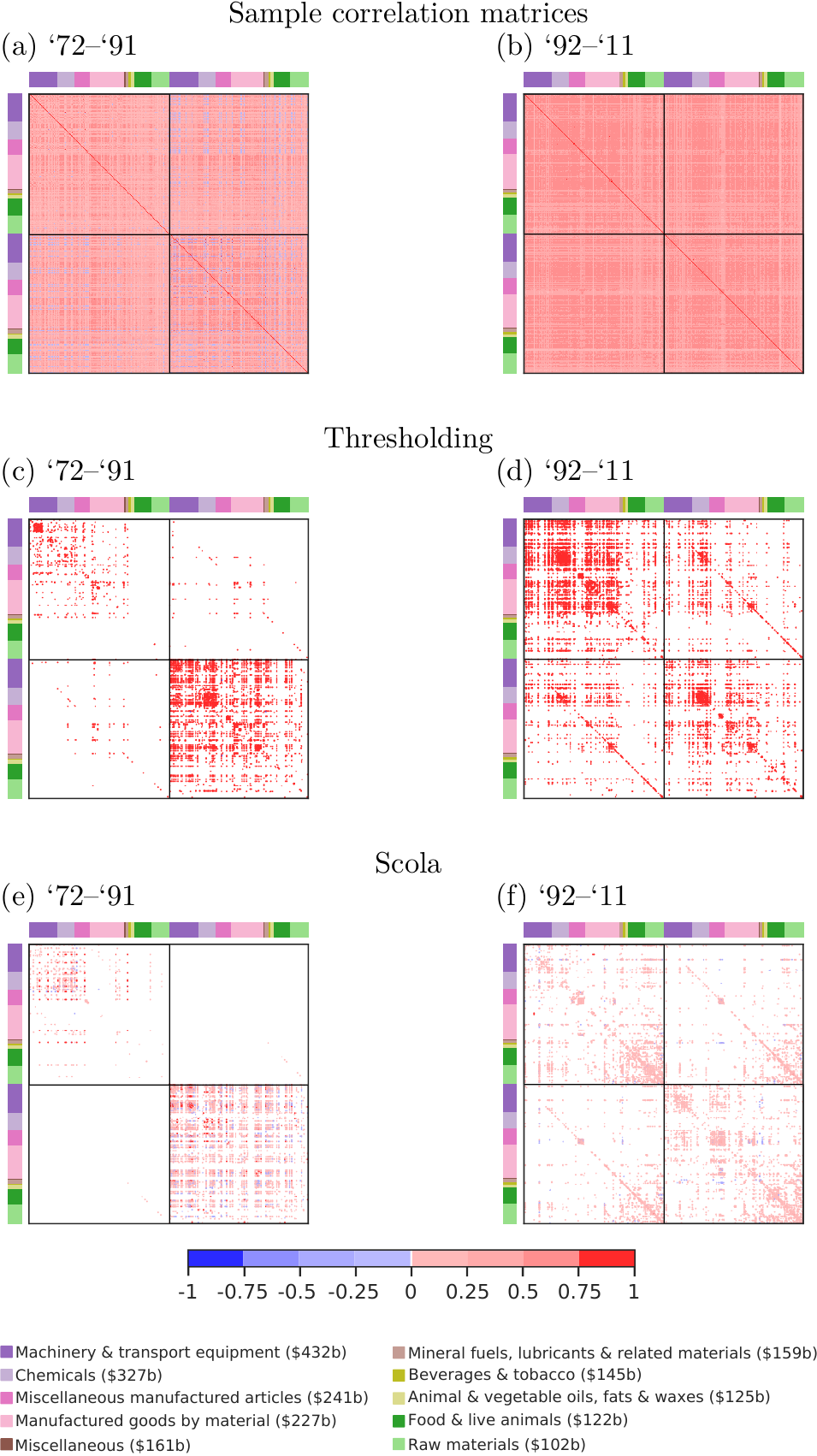}
\caption{
Sample correlation matrices and networks for the product space data.
The solid lines inside the matrices indicate the boundary between year $t$ (first half) and year $t+10$ (second half). 
The node colour indicates the product type. 
The value of the PRODY index averaged over the nodes in the same type is shown in the parentheses.
}
\label{fig:net-product-space}
\end{figure}

The sample correlation matrices are shown in Figs.~\ref{fig:net-product-space}(a) and \ref{fig:net-product-space}(b).
The thresholding method places a majority of edges in the on-diagonal blocks for both time windows (Figs.~\ref{fig:net-product-space}(c) and (d)), suggesting strong correlations between the products exported within the same year as compared to those between different years. 
This is also true for the networks generated by the Scola (Figs.~\ref{fig:net-product-space}(e) and (f)).

\begin{figure}
\centering
\includegraphics[width=0.9\textwidth]{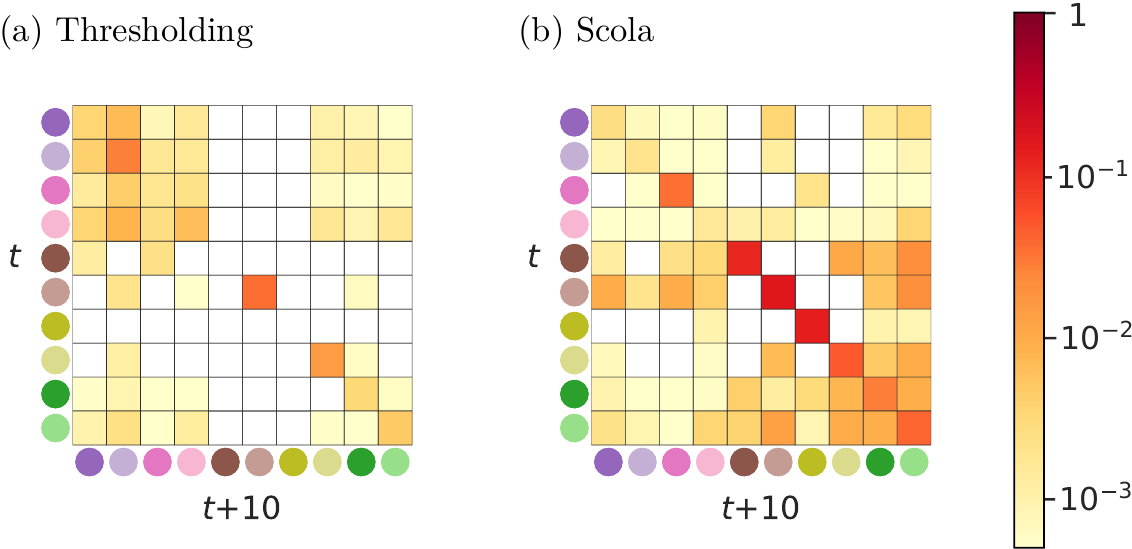}
\caption{
    Fraction of edges between product types for time window `92--`11.
}
\label{fig:product-space-mnet}
\end{figure}

The two methods place few edges (less than 5\%) within the off-diagonal blocks for `72--`91.
For `92--`11, more than 22\% of edges are placed within the off-diagonal blocks by both methods. 
Although both methods place a similar number of edges within the off-diagonal blocks, the distribution of edges is considerably different.
To see this, we compute the fraction of edges between product types for `92--`11 (Fig.~\ref{fig:product-space-mnet}).
We do not show the result for `72--`91 owing to a small number of edges within the off-diagonal blocks.  
We find that, within the off-diagonal blocks, the thresholding method places relatively many edges between two nodes that correspond to sophisticated products in terms of the PRODY index (Fig.~\ref{fig:product-space-mnet}(a)).
Examples include ``Machinery \& transport equipment'', ``Chemicals'' and ``Manufactured goods by material''. 
This result suggests that sophisticated products are strongly correlated with the same or other sophisticated products ten years apart.  
In contrast, the Scola finds many edges between nodes that correspond to less sophisticated products such as ``Raw materials'', ``Foods \& live animals''  and ``Animal \& vegetable oils, fats \& waxes'' (Fig.~\ref{fig:product-space-mnet} (b)).
We find a similar result for the on-diagonal blocks for `92--`11 (Fig.~\ref{fig:net-product-space}(f)).

\begin{figure}
\centering
\includegraphics[width=\hsize]{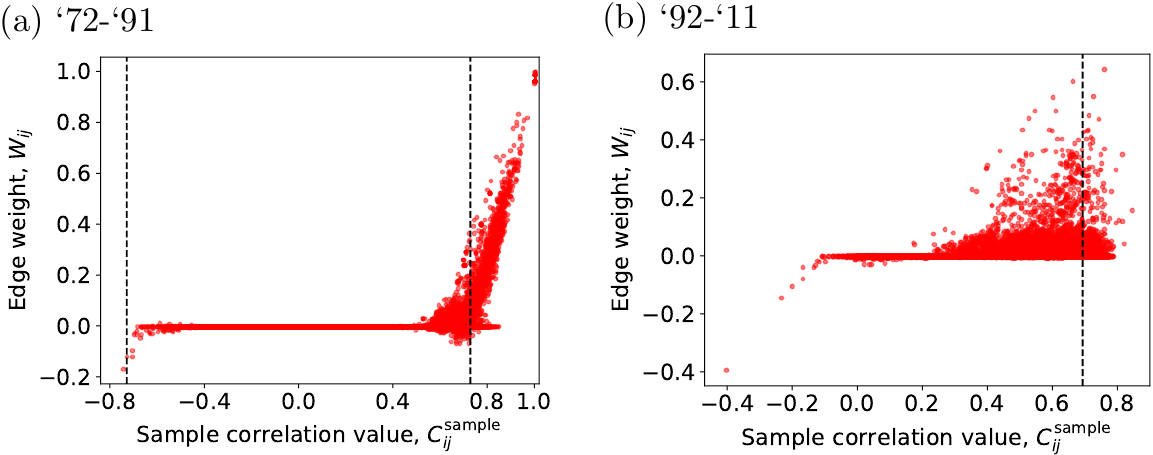}
\caption{
Correlation values in the sample correlation matrix and the weight of edges in the networks generated by the Scola shown in Figs.~\ref{fig:net-product-space} (e) and (f).  
The dashed lines indicate the threshold value adopted by the thresholding method. 
}
\label{fig:corr-vs-weight-product-sapce}
\end{figure}

Highly correlated nodes may not be adjacent in the network generated by the Scola. 
To examine this issue, we plot the weight of edges estimated by the Scola against the correlation value of the corresponding node pair in Fig.~\ref{fig:corr-vs-weight-product-sapce}.
For time window `72--`91, there are 3,021 node pairs with a correlation above the threshold in magnitude, of which 1,700 (56\%) node pairs are not adjacent in the network generated by the Scola.
We find qualitatively the same result for time window `92--`11; there are 6,834 node pairs with a correlation above the threshold in magnitude, of which 6,343 node pairs ($93\%$) are not adjacent in the network generated by the Scola.
The weights of edges are correlated strongly with the original correlation values for time window `72--`91 but weakly for `92--`11 (Spearman correlation coefficients are 0.87 and 0.21, respectively).

We further demonstrate the use of the generated networks for predictions. 
We aim to predict RCA values in year $t+10$ given those in year $t$. 
We make predictions using vector autoregression \cite{Kilian2006}, which consists in computing conditional probability distribution $P(\xvec_{(c,t+10)}|\xvec_{(c,t)}) = P(\xvec_{(c,t)},\xvec_{(c,t+10)})/ P(\xvec_{(c,t)})$. Note that the joint probability distribution $P(\xvec_{(c,t)}, \xvec_{(c,t+10)})$ is given by Eq.~\eqref{eq:likelihood}.
The joint probability distribution $P(\xvec_{(c,t)}, \xvec_{(c,t+10)})$ has covariance matrix $\Cov$ as parameter.
We set $\Cov$ to either the sample correlation matrix ($\Cov = \sCov$) or that provided by the Scola (i.e., $\Cov=\nCov +\qCov$).
We do not use the thresholding method in this prediction task because the thresholding method does not provide correlation matrices.
We make a prediction by the conditional expected value of $\xvec_{(c,t+10)}$ under $P(\xvec_{(c,t+10)}|\xvec_{(c,t)})$, which is given by \cite{Bishop2006}
\begin{align}
	\Exp[\xvec_{(c,t+10) } \vert \xvec_{(c,t) } ] &= \int \xvec_{(c,t+10) } P\left(\xvec_{(c,t+10)}|\xvec_{(c,t)}\right) {\rm d}\xvec_{(c,t+10)}  \nonumber \\
	&=  \xvec_{(c,t)}\Cov^{-1} _{0, 0}\Cov _{0, +10}, 
\end{align}
where $\Cov _{0, 0}$ and $\Cov_{0, +10}$ are the blocks of $\Cov$ defined analogously to Eq.~\eqref{eq:scov-block}.

We carry out five-fold cross-validation, where we split sample indices $\{1,\ldots, L\}$ into five subsets of almost equal sizes.
We estimate $\Cov$ using the training set, which is the union of four out of the five subsets. 
Then, we perform predictions for the test set, which is the remaining subset.
We carry out this procedure five times such that each of the subsets is used once as the test set.

The joint distribution of the actual and predicted $\overline \RCA_{cp} ^{(t+10)}$ values, where each tuple $(c,p,t)$ is regarded as an individual data point, is shown in Fig.~\ref{fig:product-space-prediction}.
(The joint distributions for other null models are shown in the Supplementary Materials.)
The perfect prediction would yield all points on the diagonal line. 
Between `72--`91, the Scola realises better predictions than the sample correlation matrix does (Figs.~\ref{fig:product-space-prediction}(a) and \ref{fig:product-space-prediction}(c)).
In fact, the mean squared error (MSE) is approximately three times smaller for the Scola than for the sample correlation matrix. 
Between `92--`11, the MSE for the Scola is approximately 1.5 times smaller than that for the sample correlation matrix (Figs.~\ref{fig:product-space-prediction}(b) and \ref{fig:product-space-prediction}(d)). 
A probable reason for the poor prediction performance for the sample correlation matrices is overfitting.
Matrix $\sCov$ contains more than $1.9\times 10^6$ elements, whereas there are only  $L=$1,400 samples on average.
The Scola represents the correlation matrix with a considerably smaller number of parameters (i.e., $M+K=$8,185 on average), which mitigates overfitting.

\begin{figure}
\centering
\includegraphics[width=\hsize]{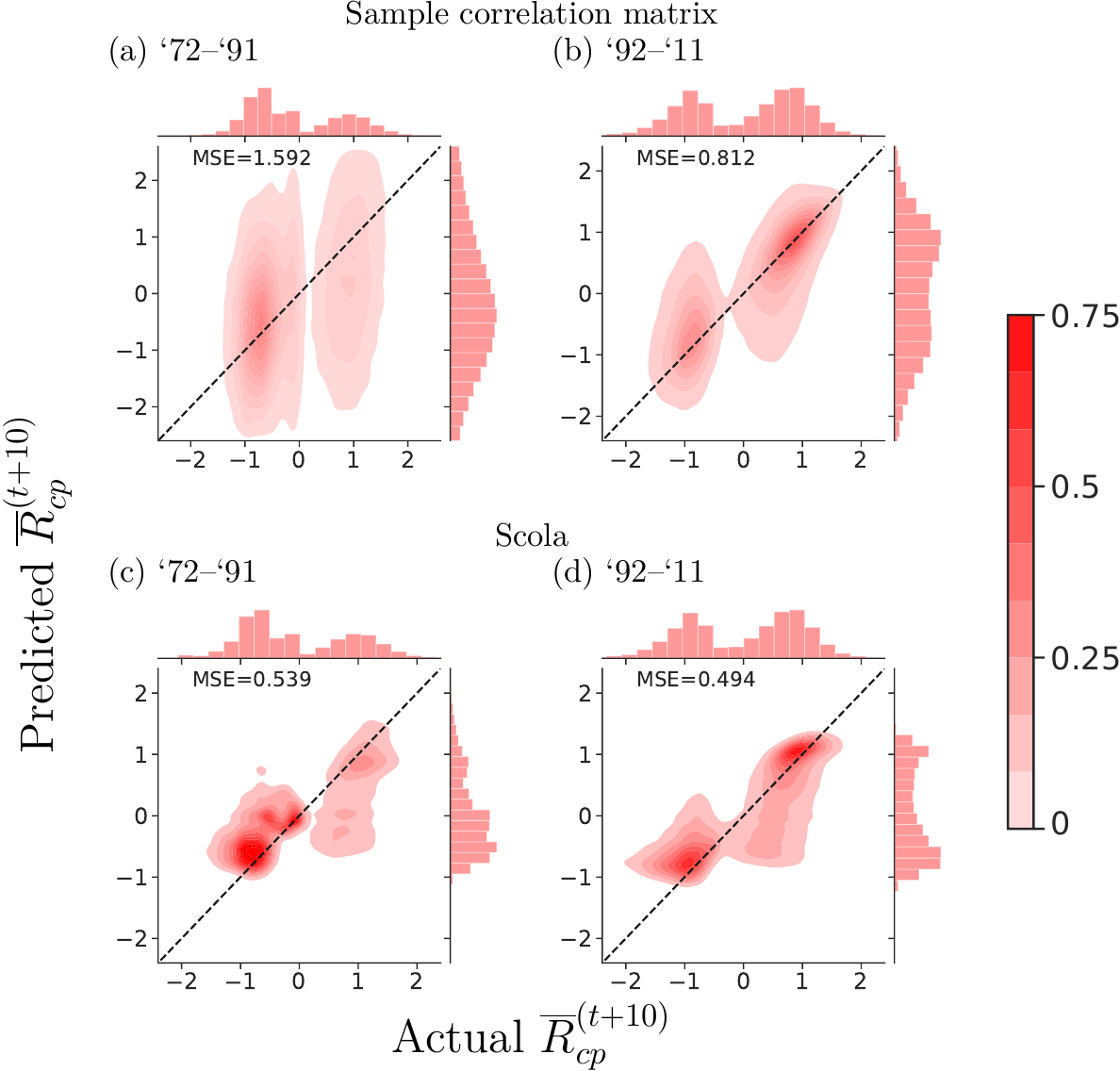}
\caption{
Prediction of product exports.
Joint distributions of the actual and predicted RCA values are shown.
The dashed lines represent the diagonal. 
The MSE represents the mean square error. 
The marginal distributions are shown to the top and right of each panel.
}
\label{fig:product-space-prediction}
\end{figure}
 
\subsection{Model selection}
\label{sec:model-selection-result}

What are appropriate null models for correlation networks? 
To address this question, we carry out model selection based on the EBIC to compare the Scola with different null models. 
We examine the three null models, i.e., the white noise model, the HQS model and the configuration model. 
We also compare the performance of the Scola with estimators of sparse precision matrices with different null models.

To construct a network from a precision matrix, we use a variant of the Scola (Appendix~\ref{sec:precisionmatrix}).
We adopt the inverse of the correlation matrices for the white noise model, the HQS model and the configuration model as the null models for precision matrices.
Because the white noise model is the identity matrix, its inverse is also the identity matrix, providing the white noise model for precision matrices. 
It should be noted that the variant is equivalent to the graphical Lasso \cite{Banerjee2008,Friedman2008} when one uses the white noise model as the null model.

In addition to the country-level export data used in Section~\ref{sec:product-space}, we use the time series of stock prices in the US and Japanese markets.
The US data comprise the time series of the daily closing prices of $N=$1,023 companies in the list of the Standard \& Poor's 500 (S\&P 500) index on 4,174 days between 01/01/2000 and 29/12/2015.
We compute the logarithmic return of the daily closing price by $x_{ti} \equiv \log\left[z_i(t+1) / z_i(t)\right]$, where $z_i(t)$ is the closing price of stock $i$ on day $t$.
We split the time series of $x_{ti}$ into two halves of eight years.
Then, for each half, we exclude the stocks that have at least one missing value. 
(The stock prices are missing at the time points at which companies are not included in S\&P 500.)
We compute the correlation matrix for the logarithmic returns of the remaining stocks for each half.

The Japanese data comprise the time series of the daily closing prices of 264 stocks in the first section of the Tokyo Stock Exchange.
We retrieve the stock prices for 5,324 days between 12/03/1996 and 29/02/2016 using Nikkei NEED \cite{NIKKEI}, where we exclude the stocks that do not have transactions on at least one day during the period.
As is the case for the US data, we compute the correlation matrix for the logarithmic returns.
We used the same correlation matrix for the Japanese data in our previous study \cite{Masuda2018}.
We refer to the US and Japanese stock data as S\&P 500 and Nikkei, respectively. 

The EBIC values for the networks generated by the Scola and its variant for precision matrices combined with the three null models are shown in Table~\ref{ta:bic_comparison}. 
For the product space and S\&P 500, the EBIC value for the configuration model is the smallest.
For Nikkei, the EBIC value for the HQS model for precision matrices is the smallest.
This result suggests that the Scola does not always outperform its variant for precision matrices.
It should be noted that the graphical Lasso, which is equivalent to the variant of the Scola for precision matrices combined with the white noise model, is among the poorest across the different data sets.

\begin{table}
\centering
\caption{
Normalised EBIC values for correlation matrices and precision matrices combined with the different null models.
We divide the EBIC value for each null model by that for the sample correlation matrix.
Matrices $\Cov^{\text{WN}}$,  $\Cov^{\text{HQS}}$ and  $\Cov^{\text{con}}$ indicate the white noise model, HQS model and configuration model, respectively, as null models.
For null precision matrices, we adopt the inverse of the three null correlation matrices and construct networks using a variant of the Scola (Appendix~\ref{sec:precisionmatrix}).
}
\label{ta:bic_comparison}
\scalebox{1.00}{
\begin{tabular}{ccccccccc}
\toprule
    && \multicolumn{7}{c}{Null model}\\ \cline{3-9}
    \multirow{2}{*}{Data} && \multicolumn{3}{c}{Correlation matrix} & & \multicolumn{3}{c}{Precision matrix} \\ 
    && $\Cov^{\text{WN}}$ & $\Cov^{\text{HQS}}$ & $\Cov^{\text{con}}$  && $\Cov^{\text{WN}}$ & $\Cov^{\text{HQS}}$ & $\Cov^{\text{con}}$  \\
\hline
    \multicolumn{5}{l}{Product space}  \\ 
            `72--`91 & & 0.226 & 0.202 & 0.187 && 0.232 & 0.190 & 0.208 \\
            `92--`11 & & 0.300 & 0.301 & 0.226 && 0.304 & 0.238 & 0.256 \\ \\
    \multicolumn{5}{l}{S\&P 500} \\ 
            `00--`07 & &  0.600 & 0.562 & 0.553 && 0.640 & 0.557 & 0.587 \\
            `08--`15 & &  0.664 & 0.558 & 0.524 && 0.596 & 0.539 & 0.607 \\ \\
    \multicolumn{1}{l}{Nikkei}  &&  0.991 & 0.874 & 0.861 && 1.001 & 0.837 & 0.882 \\
\botrule
\end{tabular}
}
\end{table}

\section{Discussion}
\label{sec:discussion}

We presented the Scola to construct networks from correlation matrices.
We defined two nodes to be adjacent if the correlation between them is significantly different from that expected for a null model.
The Scola yielded insights that were not revealed by the thresholding method such as a positive correlation between less sophisticated products across a decade.
The generated networks also better predicted country-level product exports after ten years than the mere sample correlation matrices.

Null models that have to be fed to the Scola are not limited to the three models introduced in Section~\ref{sec:null-models}. 
Another major family of null models for correlation matrices is those based on random matrix theory, which preserves a part of spectral properties of given correlation matrices \cite{Vasiliki2002,Utsugi2003,MacMahon2015}. 
The Scola cannot employ this family of null models because it requires the null correlation matrix to be invertible.
Null matrices based on random matrix theory are often not invertible because they leave out some of the eigenmodes.
A remedy to this problem is to use a pseudo inverse.

The Scola is equivalent to the previous algorithm \cite{Bien2011} if one adopts the white noise model as the null model, i.e., $\nCov = \Cov^{\text{WN}}$. 
Another method closely related to the Scola is the graphical Lasso \cite{Banerjee2008,Friedman2008,Fan2009,Foygel2010,VanBorkulo2014}, which provides sparse precision matrices (i.e., $\Cov^{-1}$).
In contrast to correlation matrices, precision matrices indicate correlations between nodes with the effect of other nodes being removed. 
We focused on correlation matrices because null models for correlation matrices are relatively well studied \cite{Vasiliki2002,Utsugi2003,Hirschberger2007,MacMahon2015,Masuda2018}, while studies on null models for precision matrices are still absent to the best of our knowledge.

The inverse of the correlation matrices provided by the HQS model and the configuration model may be reasonable null models for precision matrices.
To explore this direction, we developed a variant of the Scola for the case of precision matrices, which is equivalent to the graphical Lasso if the white noise model is the null model.
The variant of the Scola generates networks better than the original Scola for the Nikkei data in terms of the EBIC (Section \ref{sec:model-selection-result}).
We do not claim that the Scola is generally a strong performer. More comprehensive comparisons of the Scola and competitors warrant future work.

Although the thresholding method has been widely employed \cite{Kose2001,Onnela2004,Rubinov2010,VanWijk2010,Zanin2012,DeVicoFallani2014}, the overfitting problem inherent in this method has received much less attention than it deserves. 
In many cases, the number of observations based on which one computes the correlation matrix is of the same order of the number of nodes, which is much smaller than the number of the entries in the correlation matrix \cite{Banerjee2008,Bien2011}. 
In this overfitting situation, if one removes or adds a small number of observations, one may obtain a substantially different correlation matrix and the resulting network.
 
We have assumed that the input data obey a multivariate Gaussian distribution.
However, this assumption may not hold true for empirical data. 
A remedy commonly used in machine learning is to transform data using an exponential function, which is referred to as power transformations.
The Box-Cox transformation, which we used in the analysis of the product space, is a standard power transformation.   
Other transformation techniques include the Fisher transformation \cite{Fisher1915} and the inverse hyperbolic transformation \cite{Borbidge1988}. 
Alternatively, one may assume other probability distributions for input data, as with a graphical Lasso for binary data \cite{VanBorkulo2014}. 

Although we illustrated the Scola on economic data, the method can be applied to correlation data in various fields including neuroimaging \cite{Rubinov2010,VanWijk2010,Zalesky2012,Zanin2012,DeVicoFallani2014}, psychology \cite{VanBorkulo2014}, climate \cite{Tsonis2006}, metabolomics \cite{Kose2001} and genomics \cite{Delafuente2002}. 
For example, in neuroscience, the correlation data are often used to construct functional brain networks, analysis of which is expected to provide insights into how brains operate and cognition occurs.
Applications of the Scola with the aim of finding insights that have not been obtained by thresholding methods, which has conventionally been used for these data, warrant future work.

%
%
%
%
\appendix

\section*{Acknowledgments}
The Standard \& Poor's 500 data were provided by CheckRisk LLP in the UK.
We thank Yukie Sano for providing the Japanese stock data used in the present paper. 
N.~M. acknowledges the support provided through JST, CREST, Grant Number JPMJCR1304.

\section*{Competing interests}
The authors declare no competing interests.

\section{Golden-section search}
\label{sec:golden-section-search}

We adopt the golden-section search to find the $\overline \lambda$ value that minimises the EBIC \cite{Press2007}.
Let $\overline \lambda^*$ be the $\overline \lambda$ value yielding the minimum EBIC value. 
Let $\EBIC(\overline \lambda)$ be the EBIC value at $\overline \lambda$.
In most cases, when one increases $\overline \lambda$ from 0, the EBIC monotonically decreases for $\overline \lambda \leq \overline \lambda^*$, reaches the minimum at $\overline \lambda = \overline \lambda^*$ and monotonically increases for $\overline \lambda > \overline \lambda^*$.
The golden-section search method exploits this property to find $\overline \lambda^*$.
Specifically, suppose that one knows a lower bound and an upper bound of $\overline \lambda^*$, i.e., $\overline \lambda_{\text{lower}} \leq \overline \lambda^* \leq \overline \lambda_{\text{upper}}$.
Then, consider $\overline \lambda_1$ and $\overline \lambda_2$, where $\overline \lambda_{\text{lower}} < \overline \lambda_1 < \overline \lambda_2 < \overline \lambda_{\text{upper}}$.
If $\EBIC(\overline \lambda_1) < \EBIC(\overline \lambda_2)$, it indicates $\overline \lambda^* \leq \overline \lambda_2$, yielding a tighter bound $\overline \lambda^* \in [\overline \lambda_{\text{lower}}, \overline \lambda _{2}]$.
In contrast, $\EBIC(\overline \lambda_1) > \EBIC(\overline \lambda_2)$ indicates $\overline \lambda^* \geq \overline \lambda_1$, yielding $\overline \lambda^* \in [\overline \lambda_{1}, \overline \lambda _{\text{upper}}]$.

Based on this observation, the golden-section search method iterates the following rounds. 
In round $k=0$, we set the initial lower and upper bounds following a previous study on regression analysis with Lasso \cite{Friedman2010}.
Specifically, we set $\overline \lambda^{(1)} _{\text{lower}} = 0$ and $\overline \lambda_{\text{upper}}^{(1)}$ to the minimum value of $\overline \lambda$ that satisfies ${\cal S}_{\bm \lambda}\left(\partial F_{\overline \qCov} (\overline \qCov)/\partial \overline \qCov\right)={\bm 0}$ at $\overline \qCov={\bm 0}$, i.e.,
\begin{align}
    \overline \lambda_{\text{upper}}^{(1)} = \max_{\substack{i,j\\i>j}} \left| \left[\left(\nCov\right)^{-1} - \left(\nCov\right)^{-1}\sCov\left(\nCov\right)^{-1}\right]_{ij} \times \left(W_{ij} ^{\text{MLE}}\right)^2  \right|.
\end{align}
In round $k\geq 1$, one sets $\overline \lambda_1 ^{(k)}= \overline \lambda_{\text{lower}} ^{(k)}+ h^{(k)}/\phi^2$ and $\overline \lambda_2  ^{(k)}= \overline \lambda_{\text{lower}} ^{(k)}+ h^{(k)}/\phi$, where $h^{(k)} \equiv \overline \lambda^{(k)}_{\text{upper}}-\overline \lambda^{(k)}_{\text{lower}}$ and $\phi = (1 + \sqrt{5})/2$ is the golden ratio.
Then, one updates the bound, i.e.,  
$(\overline \lambda^{(k+1)}_{\text{lower}}, \overline \lambda^{(k+1)}_{\text{upper}}) = (\overline \lambda_{\text{lower}} ^{(k)}, \overline \lambda^{(k)} _{2})$ if $\EBIC(\overline \lambda_1 ^{(k)}) < \EBIC(\overline \lambda_2 ^{(k)})$ and $(\overline \lambda^{(k+1)}_{\text{lower}}, \overline \lambda^{(k+1)}_{\text{upper}}) = (\overline \lambda_{1} ^{(k)}, \overline \lambda^{(k)} _{\text{upper}})$ if $\EBIC(\overline \lambda_1 ^{(k)}) > \EBIC(\overline \lambda_2 ^{(k)})$. 
If $\EBIC(\overline \lambda_1 ^{(k)}) = \EBIC(\overline \lambda_2 ^{(k)})$, we adopt either bound pair with an equal probability. 
If $h^{(k+1)}< h^{(1)}/100$, we stop the rounds of iteration and take the smaller of $\EBIC(\overline \lambda_{\text{lower}} ^{(k+1)})$ or $\EBIC(\overline \lambda^{(k+1)} _{\text{upper}})$ as the output. 
Otherwise, we carry out the $(k+1)$th round.

\section{Preprocessing RCA values}
\label{sec:box-cox}

The distribution of RCA may be considerably skewed.
To make the distribution closer to a normal distribution, we perform a Box-Cox transformation \cite{Box1964}, i.e., $\log\left(\RCA ^{(t)} _{cp} + 10^{-8}\right)$.
Then, for each product, we compute the $z$-score for the transformed values, generating normalised samples $\overline \RCA ^{(t)} _{cp}$  that have average zero and variance one for each product. 

For the prediction task, we transform the RCA values separately for each time window as follows.
First, we perform the Box-Cox transformation. 
Then, we split the transformed RCA values into a training set and test set.
We compute the $z$-score for the training samples, generating normalised samples.
For the test samples, we compute $z$-score for each product using the average and variance for the {\it training} samples because we should assume that 
statistics of the test samples are unknown when predicting the values of the test samples.

\section{Constructing networks from precision matrices}
\label{sec:precisionmatrix}

The precision matrices are the inverse of the correlation matrix and contain the correlation between nodes with the effect of other nodes being removed. 
The precision matrix is sensitive to noise in data, which calls for robust estimators such as the graphical Lasso \cite{Banerjee2008,Friedman2008,Fan2009,Foygel2010,VanBorkulo2014}.
The graphical Lasso implicitly assumes a null model, where every node is conditionally independent of each other, as is the case for the white noise model for correlation matrices.
Other null models for precision matrices have not been proposed to the best of our knowledge.
Nevertheless, one may use the inverse of the null models for correlation matrices as the null models for precision matrices.  

Therefore, we develop a variant of the Scola for precision matrices as follows.
Denote by $\Pre$ the precision matrix (i.e., $\Pre = \Cov^{-1}$).
We write the precision matrix as 
\begin{align}
	\Pre\equiv \nPre + \qPre,
\end{align}
where $\nPre$ is the null precision matrix. 
Note that we have redefined $\qPre$ by the deviation of the precision matrix from the null precision matrix $\nPre$.
The penalized log likelihood function (Eq.~\eqref{eq:plikelihood}) is rewritten as 
\begin{align}
	\hat {\cal L}
		\left(\qPre \vert {\bm \lambda} \right) &=  
		-\frac{L}{2}\ln\det\left(\nPre + \qPre \right)^{-1} - \frac{L}{2}\trace\left[ \sCov(\nPre + \qPre)\right]  - \frac{NL}{2} \ln(2\pi)\nonumber \\
		&- \frac{L}{2}\sum_{i=1}^N \sum_{j=1}^{i-1} \lambda_{ij} |\qpre{ij}|. 
\end{align}
We note that $\hat {\cal L}$ is a concave function with respect to $\qPre$, which is different from the case of correlation matrices.
By exploiting the concavity, we maximise $\hat {\cal L}$ using a gradient descent algorithm instead of the MM algorithm.  
Specifically, starting from an initial solution $\widetilde \qPre^{(0)}={\bm 0}$, we update tentative solution $\widetilde \qPre^{(k)}$ until convergence.
The update equation is given by
\begin{align}
\widetilde \qPre^{(k+1)} &= {\cal S}_{\epsilon {\bm \lambda}}\left( \tilde \qPre^{(k)} - 
		\epsilon \left[ \left( \nPre + \widetilde \qPre^{(k)} \right)^{-1} - \sCov  \right]\right).
\end{align}
Each update requires the inversion of an $N\times N$ matrix, resulting in time complexity of the entire algorithm of ${\cal O}(N^3)$. 
One may be able to use conventional efficient optimisation algorithms for the graphical Lasso to save time \cite{Friedman2008}.
However, we do not explore this direction.

\includepdf[pages={{},1,2,3}]{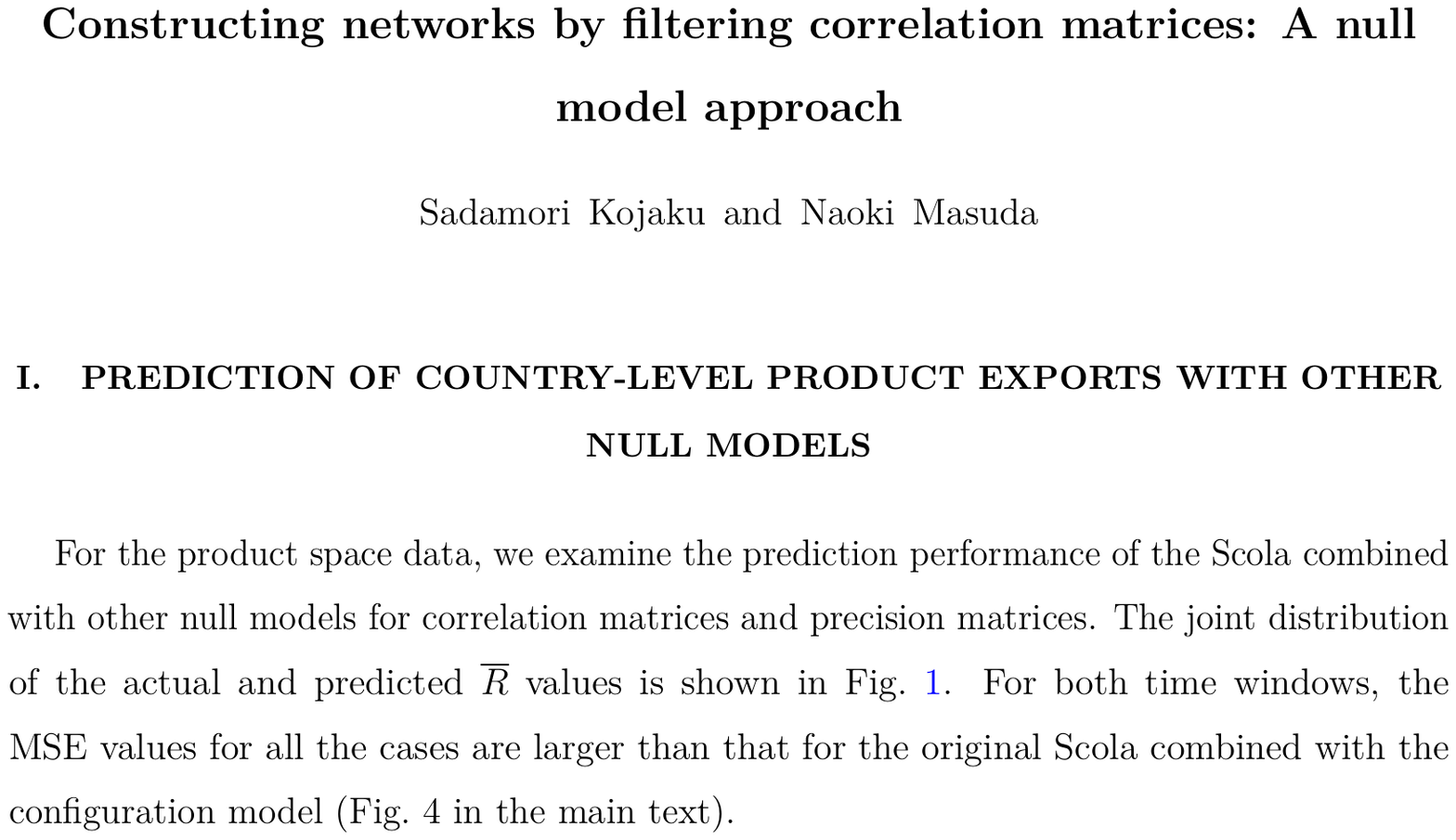}
\end{document}